\documentclass[useAMS,usenatbib,usegraphicx]{mn2e}
\title[Intergalactic dust]
{Constraint on intergalactic dust from thermal history of
intergalactic medium}
\author[A. K. Inoue \& H. Kamaya]
{Akio K. Inoue\thanks{E-mail:inoue@kusastro.kyoto-u.ac.jp}\thanks{
Research Fellow of the JSPS}  and Hideyuki Kamaya\\ 
Department of Astronomy, Faculty of Science, Kyoto University,
Sakyo-ku, Kyoto 606-8502, Japan}
\begin{document}

\date{Submitted on 2002 December 5}

\pagerange{\pageref{firstpage}--\pageref{lastpage}} \pubyear{2003}

\maketitle

\label{firstpage}

\begin{abstract}

This Letter investigates the amount of dust in the intergalactic medium 
(IGM). The dust photoelectric heating can be the most efficient heating 
mechanism in the IGM where the density is very small and there are 
a lot of hard ultraviolet photons.
Comparing the observational thermal history of IGM with a theoretical 
one taking into account the dust photoelectric heating, 
we can put an upper limit on the dust-to-gas ratio, ${\cal D}$, in the IGM.  
Since the rate of the dust photoelectric heating depends on
the size of dust, we find the following results: 
If the grain size is $\ga 100$ \AA, 
${\cal D}$ at $z \sim 3$ is $\la 1/100$ Galactic value corresponding to
$\Omega_{\rm dust}^{\rm IGM}\la 10^{-5}$.
On the other hand, if the grain size is as small as $\sim 10$ \AA, 
${\cal D}$ is $\la 1/1000$ Galactic  value corresponding to
$\Omega_{\rm dust}^{\rm IGM}\la 10^{-6}$.  

\end{abstract}


\section{Introduction}

There are metals in the intergalactic medium (IGM) even in low density
regions such as Lyman $\alpha$ forest (e.g., \citealt{cow95,tel02}).
Since metal and dust are relating each other, it is sure that dust
grains also exist in the IGM. However, the amount of the intergalactic
(IG) dust is still quite uncertain although it seems to be not so abundant.  
We try to put a new constraint on the amount of the IG dust 
by using the IGM thermal history suggested by the recent
observations of Lyman $\alpha$ forest (e.g., \citealt{sch00}). 

After the early attempts to estimate the IG extinction
\citep{eig49,hum56}, \cite{cra73} have obtained an upper limit on the
amount of the IG dust in terms of the density parameter as $\Omega_{\rm
dust}^{\rm IGM}(z=0) \la 10^{-4}$, by comparing the observed spectral
energy distribution of distant (but $z\la 0.5$) giant elliptical
galaxies with an average spectrum of similar nearby galaxies (see also
\citealt{nic71,tak72}).

Observations of the redshift evolution of the QSOs spectral slope
provide us with another approach to the IG dust \citep{wri81,che91}.  
Although Cheng et al.\ (1991) found no evidence for an appreciable IG
reddening up to $z\sim 2$--3, this may not be so strict constraint
because it is difficult to measure the UV slope of QSOs enough precisely
owing to the broad and complex iron band superposing on the continua.
Recently, we obtain further constraint on the IG dust from observations
of high-z supernovae (SNe) \citep{rie98,per99}.  An upper limit
of the colour excess by the IG reddening is $\langle E_{B-V}
\rangle_{z\sim 0.5} - \langle E_{B-V} \rangle_{z\sim 0.05} \la 0.03$
mag. 

By the way, \cite{agu99} suggests that the IG dust can be ``gray'' by a
selection rule in the transfer of grains from the host galaxies to the
IGM. This occurs when small ($\la 0.1 \micron$) grains are destroyed
selectively by the thermal sputtering in the hot gas halo of the host
galaxies before they reach the IGM. Then, the extinction property
becomes ``gray'' since the relatively large dust survives. Importantly, 
such ``gray'' IG dust can avoid the detection by the IG
reddening survey like above.

Even if the IG dust is really ``gray'', its evidence should be
imprinted in the cosmic microwave background (CMB) and infrared
background because the dust emits thermal radiation in the wave-band
from the far-infrared to submillimetre (submm)
\citep{row79,wri81}. According to the current status, {\it COBE} data
provides us with a rough upper limit on the IG dust
\citep{loe97,fer99,agu00}. On the other hand, the submm background
radiation may give a more strict constraint on the IG dust. The
emission from the IG dust expected from the modern cosmic star
formation history with the grain transfer mechanism of \cite{agu99}
contributes to a substantial fraction ($\ga$ 75\%) of the measured
background radiation in the submm \citep{agu00}. 
It is interesting to develop the approach of \cite{agu00} 
to match a strong constraint by SCUBA
whose result of the number count 
accounts $\sim$ 90\% of the submm background light (\citealt{cal01}
and references therein). 

In any current status, the amount of the IG dust is still highly uncertain.  
Hence, it is worth trying to find a new constraint on the amount of the
IG dust.  In this Letter, it is demonstrated that we can obtain such a new
constraint of the IG dust by comparing the observational IGM temperature
with a theoretical one if we take into account the dust photoelectric
heating in the theoretical model. This is possible 
because the dust heating is an
efficient mechanism in the IGM \citep{nat99}.
Interestingly, the effect of the dust photoelectric heating depends on
not only the amount but also the size of dust.
Therefore, we can put a limit on the amount of the IG dust as a
function of the typical size of the IG grains.

\section{Photoelectric Heating by Grains}


To assess photoelectric effect, we must specify the charge on grain, 
$Z_{\rm d}$ (in the electron charge unit), which is given by
\citep{spi41}:
\begin{equation}
 \sum_i R_i + R_{\rm pe}=0 \,,
 \label{eq1}
\end{equation}
where we have assumed a charge equilibrium because 
a typical charging time-scale is very short ($\sim
10[a/0.1\micron]^{-1}$ yr under the UV background radiation dominated by
QSO light given in \S 3).   
In the equation, $R_i$ is the collisional charging rate by $i$-th
charged particle, and $R_{\rm pe}$ is the photoelectric charging
rate. We assume spherical grains for simplicity.

The collisional charging rate is 
\begin{equation}
 R_i = \pi a^2 Z_i s_i n_i \int_{v_{\rm min}}^{\infty} 
  \sigma_i(v_i, Z_{\rm d}, Z_i) v_i f(v_i) dv_i \,,
 \label{eq2}
\end{equation}
where $Z_i$ is the charge in the electron charge unit, $s_i$ is the
sticking coefficient, $n_i$ is the number density, $a$ is the grain
radius, $v_i$ is the velocity, $v_{\rm min}$ is the minimum velocity
required to collide with a grain, $\sigma_i$ is the dimensionless
collisional cross section depending on both charges and $v_i$, and
$f(v_i)$ is the velocity distribution function.  We simply assume $s_i$
is always unity.  In the above equation, we neglected the effect of the
secondary emission.

We can neglect the effect of the ``image potential'' \citep{dra87} on
the collisional cross section because the obtained charges are enough
large. If we assume the Maxwellian velocity distribution for the
particle, the integral in the r.h.s.\ of equation (\ref{eq2}) is reduced
to $(8k_{\rm B}T/\pi m_i)^{1/2} g(x)$ and $g(x)=1-x$ for $Z_{\rm d}Z_i
\leq 0$ or $g(x)=\exp (-x)$ for $Z_{\rm d}Z_i > 0$, where $k_{\rm B}$ is
the Boltzmann's constant, $T$ is the gas temperature, $m_i$ is the charged
particle's mass, and $x=e^2Z_{\rm d}Z_i/ak_{\rm B}T$.

The photoelectric charging rate is given by
\begin{equation}
 R_{\rm pe}=\pi a^2 \int_{\nu_{\rm min}}^{\nu_{\rm max}} 
  Q(a,\nu) Y(a,\nu,Z_{\rm d}) \frac{4 \pi J_{\nu}}{h \nu} d\nu\,,
 \label{eq5}
\end{equation}
where $Q$ is the absorption coefficient of grains, $Y$ is the
photoelectric yield, $J_{\nu}$ is the intensity (averaged for solid
angle) of the incident radiation at a frequency $\nu$, $h$ is the Plank
constant, and $\nu_{\rm max}$ is the maximum frequency of the incident
radiation.  The minimum frequency, $\nu_{\rm min}$, is the threshold
photon frequency required for an electron to escape from a grain.  That
is, $h\nu_{\rm min}$ is equal to the ionization potential, $IP$, of a
grain: $IP=W+(Z_{\rm d}+1/2)e^2/a$, where $W$ is the work function.
Here we neglect quantum effects on $IP$ because it is very small for
grains with $a \ga 10$ \AA\ (\citealt{wei01a}, hereafter WD01).

According to WD01, we adopt $W=4.4$ eV for graphites and $W=8.0$ eV for
silicates.  For $Q$, we adopt the values of ``graphite'' and ``smoothed
UV astronomical silicate'' calculated by \citet{dra84,lao93,wei01b}. The
value of $Y$ is estimated, based on the way constructed by WD01, which
includes approximately the effect of the energy distribution of
photoelectrons and the geometrical enhancement for small grains, and
reproduces recent results of the laboratory experiments even for very
small grains.

The both equilibrium charges obtained for graphites
and silicates are similar when their radii are equal. 
This is consistent with the result of Nath et al.~(1999).
In the following discussion, we consider mainly silicate case.


The photoelectric heating rate by a grain with radius $a$ is expressed by 
\begin{equation}
 \gamma(a)= \pi a^2 
  \int_{\nu_{\rm min}}^{\nu_{\rm max}} E_{\rm pe} (a,\nu,Z_{\rm d}) 
  Q Y \frac{4 \pi J_{\nu}}{h \nu} d\nu \,,
 \label{eq7}
\end{equation}
and the cooling rate of the electron capture (recombination) by the
grain is $\lambda(a)=(3/2) k_{\rm B}T |R_{\rm e}|$, 
where we have assumed the Maxwellian distribution for the gaseous
electrons. We also have defined $E_{\rm pe}$, the mean kinetic energy of
photoelectrons that are emitted from a grain with radius $a$ and charge
$Z_{\rm d}$ when a photon with energy $h\nu$ is absorbed.  This is
determined from the energy distribution function of the photoelectrons,
$f(E)$, which is assumed to be a parabolic function introduced by WD01,
\footnote{Our $f(E)$ is equal to $f_E(E)=f^0_E(E)/y_2$ in WD01.} 
and is consistent with a typical energy distribution of photoelectrons
in the laboratory experiments. Hence, we estimate
\begin{equation}
 E_{\rm pe}=\int_0^{E_{\rm max}} Ef(E)dE
  =\frac{E_{\rm max}(E_{\rm max}-2E_{\rm min})}
  {2(E_{\rm max}-3E_{\rm min})}\,,
 \label{eq8}
\end{equation}
where $E_{\rm max}=h\nu-IP$ is the maximum energy of the photoelectrons
and $E_{\rm min}=-e^2(Z_{\rm d}+1)/a$ is the minimum energy of the
photoelectrons appearing on the grain surface (however falling back into
the grain).  For the parabolic function adopted here, a typical kinetic
energy of the photoelectrons is about half of $E_{\rm max}$.

To estimate the total photoelectric heating rate, we specify a
characteristic grain size in the IGM instead of the size distribution of
the IG grains. This is because the size distribution
in the IGM is quite unknown.  Here we consider three cases for
the typical grain size; 0.001 \micron, 0.01 \micron, and 0.1 \micron.
The smallest case of 0.001 \micron\ is based on the size of the primary
grains produced in the ejecta of SNe II \citep{tod01}.
It is interesting to obtain an information about the size of the IG dust
as well as to put limits on the amount. From the analysis of 
the case of very small IG dust, we can assess the selection rule of the
IG dust suggested by \cite{agu99}.

The grain number density is given as $n(a)=\rho_{\rm d}/m(a)$, where
$m(a)=(4\pi/3)\varrho a^3$ is the mass of a grain with the radius $a$,
$\varrho$ $(= 2\, {\rm g\,cm^{-3}})$ is the grain material density, and
$\rho_d$ is the volume grain mass density, which is given by $\rho_{\rm
d}=\mu m_{\rm H}n_{\rm b}{\cal D}$, where $\mu$ is the mean atomic 
weight, $m_{\rm H}$ is the proton mass, $n_{\rm b}$ is the baryon number
density, and ${\cal D}$ is the dust-to-gas mass ratio. We introduce a relative
dust-to-gas ratio, $\zeta$, defined as ${\cal D}=\zeta {\cal D}_{\rm MW}$,
where ${\cal D}_{\rm MW}=6\times 10^{-3}$ is the Galactic dust-to-gas 
ratio \citep{spi78}, for convenience below.  Therefore, the total
photoelectric heating rate and electron capture cooling rate per unit
volume are $\Gamma_{\rm pe}=\gamma(a) n(a)$ and $\Lambda_{\rm
pe}=\lambda(a) n(a)$, respectively.

The total heating rate is approximately proportional to $a^{-0.5}$ under
the situation considered in the next section.  
This is because the heating rate per a
grain roughly has a dependence of $a^{2.5}$, i.e., the grain cross
section plus an additional $a$ dependence caused by $\int E_{\rm pe}QY
d\nu$, whereas the grain number density is proportional to
$a^{-3}$. Therefore, for a fixed dust mass, the photoelectric effect
becomes more important as the grain size is smaller. The cooling rate
by the electron capture is always about an order of magnitude smaller
than the photoelectric heating rate.

\section{Dust and Thermal History in IGM}

Suppose an ideal fluid element with the mean density of the IGM.  The
time evolution of its temperature is governed by (e.g., \citealt{hui97}) 
\begin{equation}
 \frac{dT}{dt}=-2HT-\frac{T}{X}\frac{dX}{dt}
  +\frac{2(\Gamma-\Lambda)}{3k_{\rm B}Xn_{\rm b}}\,,
 \label{eqT}
\end{equation}
where $H$ is the Hubble constant, $n_{\rm b}$ is the cosmic mean number
density of baryon, $\Gamma$ and $\Lambda$ are the total heating and
cooling rates per unit volume, respectively, $X$ is the number ratio of
the total gaseous particles to the baryon particles, i.e., $X \equiv
\sum n_{i}/n_{\rm b}$, where $n_{i}$ is the number density of the $i$-th
gaseous species and we consider H~{\sc i}, H~{\sc ii}, He~{\sc i},
He~{\sc ii}, He~{\sc iii}, and electron.  Then, we solve equation
(\ref{eqT}) coupled with the non-equilibrium rate equations for these
gaseous particles.  For the rate coefficients, we use those compiled by
\cite{cen92}.  The adopted initial conditions are $T=T_{\rm CMB}(1+z)$,
H~{\sc i}/H(total)=0.99, He~{\sc i}/He(total)=0.99, He~{\sc
ii}/He(total)=0.0099 at redshift $z=10$, where $T_{\rm CMB}=2.7$ K is
the temperature of the CMB radiation at $z=0$. The result is insensitive
to the choice of the initial condition if calculation is started at an
epoch before H~{\sc i} reionization.  We also adopt $H_0=70$ km s$^{-1}$
Mpc$^{-1}$, $\Omega_{\rm M}=0.3$, $\Omega_\Lambda=0.7$, $\Omega_{\rm
b}=0.04$, and He mass abundance $Y=0.24$.

To add to the photoelectric heating and the electron capture cooling by
grains, we consider the following heating/cooling mechanisms: For
cooling, we use the recombination cooling, collisional
ionization/excitation cooling, bremsstrahlung cooling, and Compton
cooling compiled by \cite{cen92}.  For heating rates, to take account of
the radiative transfer effect \citep{abe99}, we multiply usual
photoionization heating rates by a correction factor of $(1+C_{\rm RT}
f_i)$, where $f_i$ is the fractional abundance of H~{\sc i}, He~{\sc i},
and He~{\sc ii} relative to total number of H and He.  This factor
mimics the situation that atoms are ionized by higher energy photons
when the optical depth is large (i.e., $f_i$ is nearly unity) relative
to the optically thin limit. 
The coefficient $C_{\rm RT}$ is a parameter determined by solving the
cosmological radiative transfer. 
The metal line cooling is negligible because we consider only the case
that metallicity is less than 1/100 solar
value and temperature is less than a few $10^4$ K (see e.g.,
\citealt{sut93}).

We need to give the UV background radiation and the cosmic
reionization history to obtain the temperature evolution.  We adopt, for
UV background, $J_\nu=J_{21} \nu^{-\alpha}$, where $J_{21}$ is the mean
intensity at the Lyman limit of H~{\sc i} in unit of $10^{-21}$ erg
s$^{-1}$ cm$^{-2}$ sr$^{-1}$ Hz$^{-1}$ and its evolution is given by
equation (3) in \cite{kit01}. We assume the spectral index to be
$\alpha=1$ as a QSO dominated background radiation.  For the
reionization history, we assume, for simplicity, a sudden reionization
of H~{\sc i} (and He~{\sc i}) at $z=6.0$
(\citealt{bec01})\footnote{Although \cite{kog03} find a higher-$z$ H~{\sc
i} reionization, our constraint of the IG dust dose not change because
it is determined mainly from the IGM temperature after the He~{\sc ii}
reionization.}; there is no UV photon before this redshift. To mimic the
He~{\sc ii} reionization at $z\approx 3.4$ \citep{the02}, we set the
maximum photon energy of the UV background to be 54.4 eV (He~{\sc ii}
Lyman limit) for $z>3.4$ and to be 1.24 keV\footnote{This is determined
by the maximum frequency in the data of grain absorption coefficient.} 
for $z \leq 3.4$ (i.e., a sudden He~{\sc ii} reionization).
Other choice of $\alpha$ will be discussed later.

We compare our theoretical thermal history at the mean density of the
IGM with the observational one obtained by \cite{sch00}. They observe
the Ly$\alpha$ forest clouds with the column density of $10^{13-15}$
cm$^{-2}$ (i.e., slightly over density regions), and then they convert
the temperature at this density range estimated from the minimum $b$
(Doppler) parameters into that at the mean density of the IGM by using
the observed equation of state of the IGM. Thus, we can compare both
thermal histories directly.

In Figure 1, we show the IGM thermal history obtained for no dust cases
and dusty cases of $\zeta=1/100$ (silicate grains). For no dust cases
(three dotted curves), the calculated thermal histories are consistent
with the data points of \cite{sch00} if $C_{\rm RT}\simeq
3.0$--7.0. For dusty cases (solid [$a=$ 0.1 \micron], dash-dotted [$a=$
0.01 \micron], and dashed curves [$a=$ 0.001 \micron]), we observe,
especially after the He~{\sc ii} reionization, temperature
enhancement by the dust photoelectric heating.  In Table 1, we
summarize the temperature enhancement factors at some redshifts for some
sets of dust size and dust-to-gas ratio.  All calculations do not include
the evolution of the dust-to-gas ratio, for simplicity.

\begin{figure}
 {\includegraphics[height=5.5cm,clip,keepaspectratio] {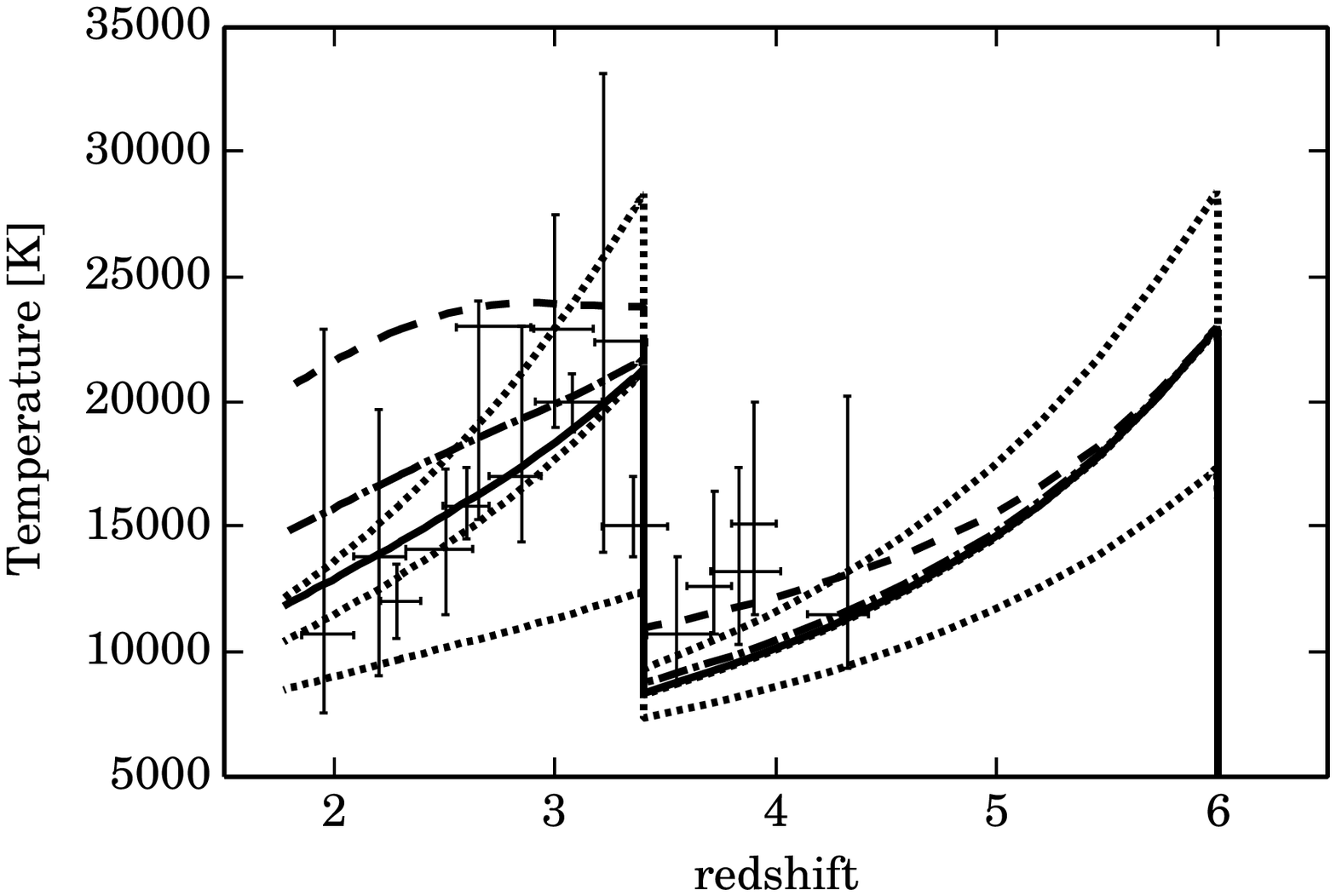}}
 \caption{IGM Thermal history under sudden reionizations
 of H~{\sc i} (and He~{\sc i}) at $z=6.0$ and He~{\sc ii} at $z=3.4$. Data
 points with errorbars are taken from Schaye et al.\ (2000). Three
 dotted curves are no dust cases; bottom: optically thin case, middle:
 $C_{\rm RT}=3.0$ case, top: $C_{\rm RT}=7.0$ case, where $C_{\rm RT}$
 is the correction factor of the radiative transfer effect. The dashed,
 dash-dotted, and solid curves are cases of grain radius $a=0.001$,
 0.01, and 0.1 \micron, respectively. For these dusty cases, the
 dust-to-gas mass ratio is set to be 1/100 Galactic value, $C_{\rm RT}=3.0$
 is assumed, and the grain type is silicate. The middle dotted curve in
 $z>3.4$ is almost superposed on the solid curve.}
\end{figure}

\begin{figure}
 {\includegraphics[height=5.5cm,clip,keepaspectratio] {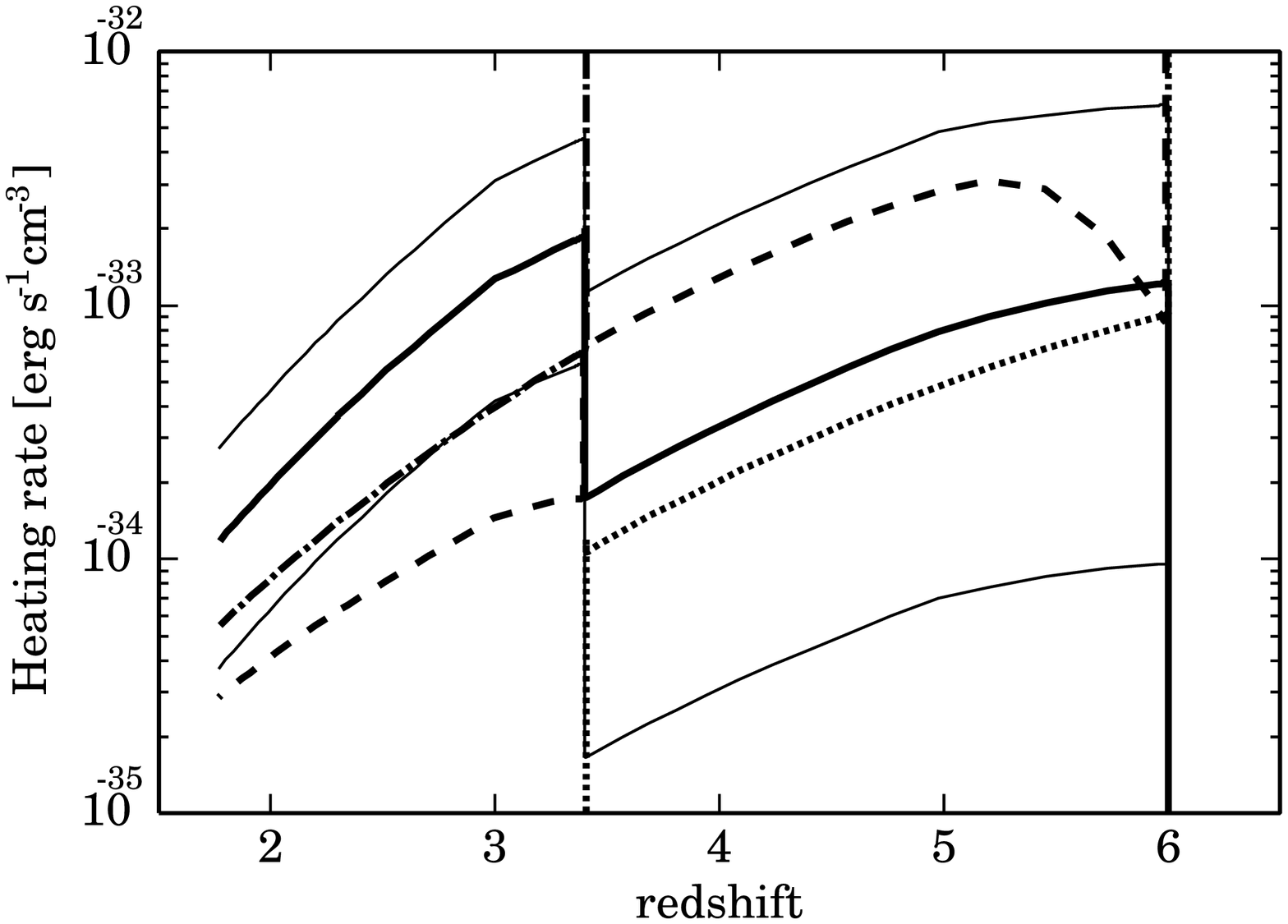}}
 \caption{Time evolution of heating rate per a unit volume for the
 case of grain radius $a=0.01$ \micron, dust-to-gas mass ratio of 1/100
 Galactic value, and radiative transfer correction $C_{\rm RT}=3.0$. 
 Solid curve is the photoelectric heating rate by grains. Dash,
 dotted, and dash-dotted curves are the photoionization heating rates by
 H~{\sc i}, He~{\sc i}, and He~{\sc ii}, respectively. Two spikes at
 $z=6.0$ and 3.4 are produced by H~{\sc i} and He~{\sc i} reionization,
 and He~{\sc ii} reionization, respectively. Two thin solid curves are
 the photoelectric heating rate of the cases of $a=0.001$ \micron\
 (top) and $a=0.1$ \micron\ (bottom). Grain type is silicate.}
\end{figure}

We also present the time evolution of heating rates per a unit volume
for the case of $a=0.01$ \micron\ and $\zeta=1/100$ in Figure 2. The
solid, dashed, dotted, and dash-dotted curves are the dust
photoelectric, H~{\sc i}, He~{\sc i}, and He~{\sc ii} photoionization
heating rates, respectively. These heating rates show sudden change at
the He {\sc ii} reionization epoch ($z=3.4$). This is because the
maximum photon energy increases from 54.4 eV to 1.24 keV suddenly
at that time. 
We note here that the heating rate by grains is saturated even when
the maximum photon energy increases more than about 300 eV in the case
of the spectral index of the incident radiation of $\alpha=1$. Thus, our
results are robust for the choice of the maximum energy after the He
{\sc ii} reionization as long as we adopt the energy larger than 300 eV.
In addition, we show the cases of $a=0.1$
and 0.001 \micron\ (and $\zeta=1/100$) by thin curves in Figure 2. Even
if we adopt other sets of $a$ and $\zeta$, the other curves of the
photoionizations do not change so significantly.

Let us discuss each case in more detail.  For the cases of $a=0.01$--0.1
\micron, before the He~{\sc ii} reionization, the photoelectric heating
by grains is minor heating mechanism relative to the H~{\sc i}
photoionization heating.  Thus, the effect of the photoelectric heating
on the IGM temperature is very small, i.e., temperature enhancement is
less than a few \%.  After the He~{\sc ii} reionization, the photoelectric
heating dominates other heating mechanisms (but by a factor of less than
3) when $\zeta=1/100$.  However, the temperature enhancement factor is
still less than 1.5 (solid and dash-dotted curve in Figure 1).  Thus,
the thermal histories of these cases are consistent with that obtained
by \cite{sch00} when $C_{\rm RT}=3.0$ is adopted.

For the case of very small ($\sim 0.001$ \micron) grains, the
photoelectric heating is always dominant when $\zeta \ga 1/100$.
Especially, this heating exceeds the He~{\sc ii} photoionization heating
by a factor of 10 after the He~{\sc ii} reionization.  Then the temperature
enhancement factor becomes about 2 at $z=2$ (dashed curve in Figure 1).
Thus, the thermal history in this case may be inconsistent with the
data from \cite{sch00}.  This means that if the typical grain size in
the IGM is $\sim 0.001$ \micron, the dust-to-gas mass ratio in the IGM
should be much less than 1/100 Galactic value.

\section{Discussion}

\begin{table}
 \centering
 \begin{minipage}{140mm}
  \caption{Temperature enhancement by IG dust.}
  \begin{tabular}{ccccc}
   \hline
   $a$, ${\cal D}/{\cal D}_{\rm MW}$ & $z=5$ & 4 & 3 & 2 \\
   \hline
   0.001 \micron, 1/1000 & 1.01 & 1.02 & 1.04 & 1.09\\
   0.001 \micron, 1/100  & 1.07 & 1.21 & 1.35 & 1.88\\
   0.01 \micron,  1/1000 & 1.00 & 1.00 & 1.01 & 1.04\\
   0.01 \micron,  1/100  & 1.01 & 1.04 & 1.13 & 1.37\\
   0.1 \micron,   1/100  & 1.00 & 1.00 & 1.04 & 1.12\\
   \hline
  \end{tabular}
 \end{minipage}
\end{table}

\begin{table}
 \centering
 \begin{minipage}{140mm}
  \caption{Amount of the IG dust at $z\ga2$.}
  \begin{tabular}{cccc}
   \hline
   $a$ & ${\cal D}/{\cal D}_{\rm MW}$ & $\Omega_{\rm dust}^{\rm IGM}$\\
   \hline
   0.001 \micron & $<10^{-3}$ & $<7\times10^{-7}$\\
   0.01--0.1 \micron   & $<10^{-2}$ & $<7\times10^{-6}$\\
   \hline
  \end{tabular}
 \end{minipage}
\end{table}

In Table 2, we summarize the obtained upper limits of the amount of the
IG dust at $z\ga2$. Let us compare these limits with the IG dust amount
estimated from a possible cosmic star formation history (e.g.,
\citealt{mpd98}). Once we assume a cosmic star formation history, the
cosmic metal amount can be estimated.  
Roughly, we find $\Omega_{\rm metal} \sim$ a few $\times 10^{-5}$ at
$z\sim3$ (e.g., \citealt{agu99}). 
If we postulate $\sim$ 0.5 as the metal depletion to dust grains,
which number is suitable for the Milky Way, we obtain $\Omega_{\rm
dust}\sim 10^{-5}$ as the total cosmic dust amount at the redshift.
We show in Table 2 the upper limits in terms
of the density parameter converted from the dust-to-gas ratios by 
$\Omega_{\rm dust}^{\rm IGM}(z)=\zeta {\cal D}_{\rm MW} \Omega_{\rm b}(z)$.
For the larger grain models, we obtain an upper limit, 
$\Omega_{\rm dust}^{\rm IGM}\sim 7\times10^{-6}$ which is
comparable to $\Omega_{\rm dust}$. This means that 
our upper limit may not deny the possibility that most of the dust
grains produced in galaxies before $z\sim 3$ escape to the IGM as an
extreme case. For very small (0.001 \micron) grain case, 
we obtain $\Omega_{\rm dust}^{\rm IGM}\la 7\times10^{-7}$.
Thus, a small fraction ($\la$ 10\%) of such very small grains can escape
from the host galaxies to the IGM 
because of $\Omega_{\rm dust}^{\rm IGM}/\Omega_{\rm dust}\la0.1$. 
The latter conclusion is consistent
with the suggestion by  \cite{agu99}, who proposes that
the IGM is very difficult to be polluted by the very small grains.

The obtained values of the heating rate by the dust photoelectric effect
are consistent with those in Nath et al.~(1999). It indicates that our
result is robust against uncertainties of the adopted efficiency factor
$Q$, photoelectric yield $Y$, and energy distribution function of the
photoelectrons, etc., because these are somewhat different between our
model and Nath et al.~(1999). If we consider only graphite grains, the
photoelectric heating rate is found to be about 70\% of that of only
silicate grains, i.e., silicate grains can be more efficient heating
source. The temperature in the graphite case is lower than that in the
silicate case, while the decrement is $\la$ 10\%. These results are  
also consistent with Nath et al.~(1999). 

We shall comment some uncertainties of our analysis.
First, the effect of the spectral index of $\alpha$ 
is examined. As shown in Zheng et al.~(1997), for example, 
a softer spectrum than $\alpha=1$ is still compatible with observations.
If $\alpha=1.5$, the heating rates by dust at $z=3$ decrease by a factor
of 0.25 ($a=$ 0.1 $\mu$m)--0.5 ($a=$ 0.001 $\mu$m) relatively to our case
of $\alpha=1$. If $\alpha=2$ is used, the heating rates at $z=3$
decrease by a factor of 0.1 (0.1 $\mu$m)--0.25 (0.001 $\mu$m).
In these calculations, $\zeta=1/100$ and $C_{\rm RT}=3.0$ are assumed. 
As the dust heating becomes less efficient, a more
abundant IG dust is allowed. Then the upper limits for the IG dust
amount increase by a factor of 2 (0.001 $\mu$m)--4 (0.1 $\mu$m) for
$\alpha=1.5$ and by a factor of 4 (0.001 $\mu$m)--10 (0.1 $\mu$m) for
$\alpha=2$ relative to the $\alpha=1$ case, because the
dust heating rate is proportional to the dust-to-gas ratio linearly. 
To summarize, a softer spectral index constrains more loosely the amount
of the IG dust by a factor.  

It might be possible to take very large $\alpha$. 
For example, we consider a background radiation field
dominated by only galaxies (e.g., $\alpha=5$). In this case,
the dust heating rates at $z=3$ become 1/100 (0.1 $\mu$m)--1/10 (0.001
$\mu$m) of those in $\alpha=1$. However, the thermal histories with
$\alpha=5$ never reproduce a steep rise of the observed
temperatures at the He~{\sc ii} reionization. In addition, the transfer
correction should be smaller as the spectrum is softer
\citep{abe99}. Therefore, the background spectrum is unlikely to be as
soft as $\alpha=5$ after the He~{\sc ii} reionization. We should assume
a harder index.

A caveat against our results is in the correction factor of the transfer
effect. Indeed, the effects of the photoelectric heating and the
transfer correction can be offset. Thus, we should check this point by
solving the cosmological radiative transfer in the future work.

The current work may give a hint for the origin of the large $b$ region
of the observed IGM clouds. There is a possibility that the large $b$
regions are localized. If the very small grains are located at the
restricted areas, a systematic heating owing to the dust photoelectrons
may explain the large-$b$ at the localized region.
The principal role of the photoelectric heating in the IGM temperature
also indicates that the theoretical equation of state in the IGM should be
reconstructed by including the photoelectric heating effect. 
All of these works are being developed by the authors.

\section*{Acknowledgments}

We would appreciate valuable comments by the referee, Dr. A.~Aguiree.
We are grateful to R.~Hirata for continuous encouragement, 
and also thank H.~Hirashita and S.~Bianchi for very useful discussion.
AKI is supported by the Research Fellowships of the
Japan Society for the Promotion of Science for Young Scientists.

\label{lastpage}
\end{document}